\documentclass[a4paper,11pt]{article}
\usepackage{pos}
\usepackage{subcaption}
\usepackage{slashed}
\usepackage{amsmath}
\allowdisplaybreaks

\newcommand{\eps}{\varepsilon}
\newcommand{\nbar}{\overline{n}}
\newcommand{\ubar}{\overline{u}}
\newcommand{\pst}{\parbox{6em}{power-suppr. terms}}

\title{Connecting Regge factorization and Collins-Soper evolution}

\author*[a]{Andrea Simonelli}

\affiliation[a]{INFN Sezione di Roma,\\
  Piazzale Aldo Moro 5, 00185, Rome, Italy}

\emailAdd{andrea.simonelli@roma1.infn.it}

\abstract{The relationship between Regge factorization of amplitudes and the Collins-Soper-Sterman factorization of hadronic cross sections is explored. Seeking a unified description of rapidity divergences in high-energy scattering and transverse-momentum-dependent factorization, a connection emerges between the Collins-Soper kernel and the Regge trajectory. This provides a framework for bridging high-energy and low-transverse-momentum resummation and for future extensions of $k_T$-factorization beyond leading-logarithmic accuracy.}

\FullConference{The 33rd International Workshop on Deep Inelastic Scattering and Related Subjects (DIS2026)\\
4 - 8 May 2026\\
Bologna, Italy\\}

\begin{document}
\maketitle

\section{Introduction}

Many scattering processes are dominated, in appropriate kinematic limits, by radiation developing along two opposite light-cone directions. 
Examples include back-to-back hadron production in $e^+e^-$ annihilation, Drell-Yan production at small transverse momentum, and inclusive deep inelastic scattering near threshold. 
Despite being distinct cases, all these processes are physical realizations of a common underlying spacetime geometry, which translates into analogous factorization properties. 
These generally involve a hard function, encoding dynamics at short distances, and two collinear functions, each associated with one light-cone direction, coupled to a soft function. 
The collinear and soft functions describe long-distance dynamics and, depending on the relevant scales, may contain genuinely non-perturbative contributions.

Partonic $2\to2$ forward scattering belongs to this broad class of processes. In this case, the two light-cone directions are naturally identified with those of the incoming partons, with momenta $p_1$ and $p_2$ aligned with the plus and minus directions, respectively. In the forward limit, the transferred momentum $l = p_3 - p_1$ is predominantly transverse and its virtuality $t = l^2 \approx -l_T^2$ is much smaller than the center-of-mass energy $s = (p_1 + p_2)^2$. 
It is then natural to ask whether the corresponding amplitude can be factorized according to the naive geometric intuition. 

\section{Forward scattering and factorization breaking}

\begin{figure}[t] 
\centering 
    \begin{subfigure}[t]{0.1\textwidth} 
    \centering 
    \includegraphics[width=\linewidth]{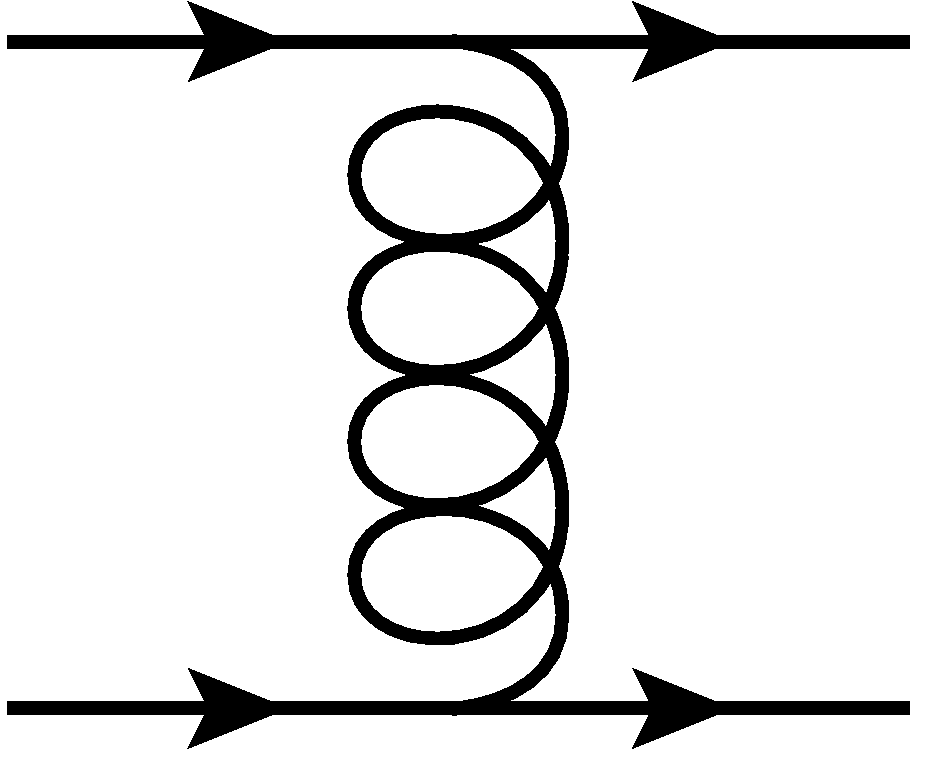} 
    \caption{} 
    \label{fig:tree} 
    \end{subfigure} 
\hspace{1cm} 
    \begin{subfigure}[t]{0.17\textwidth} 
    \centering 
    \includegraphics[width=\linewidth]{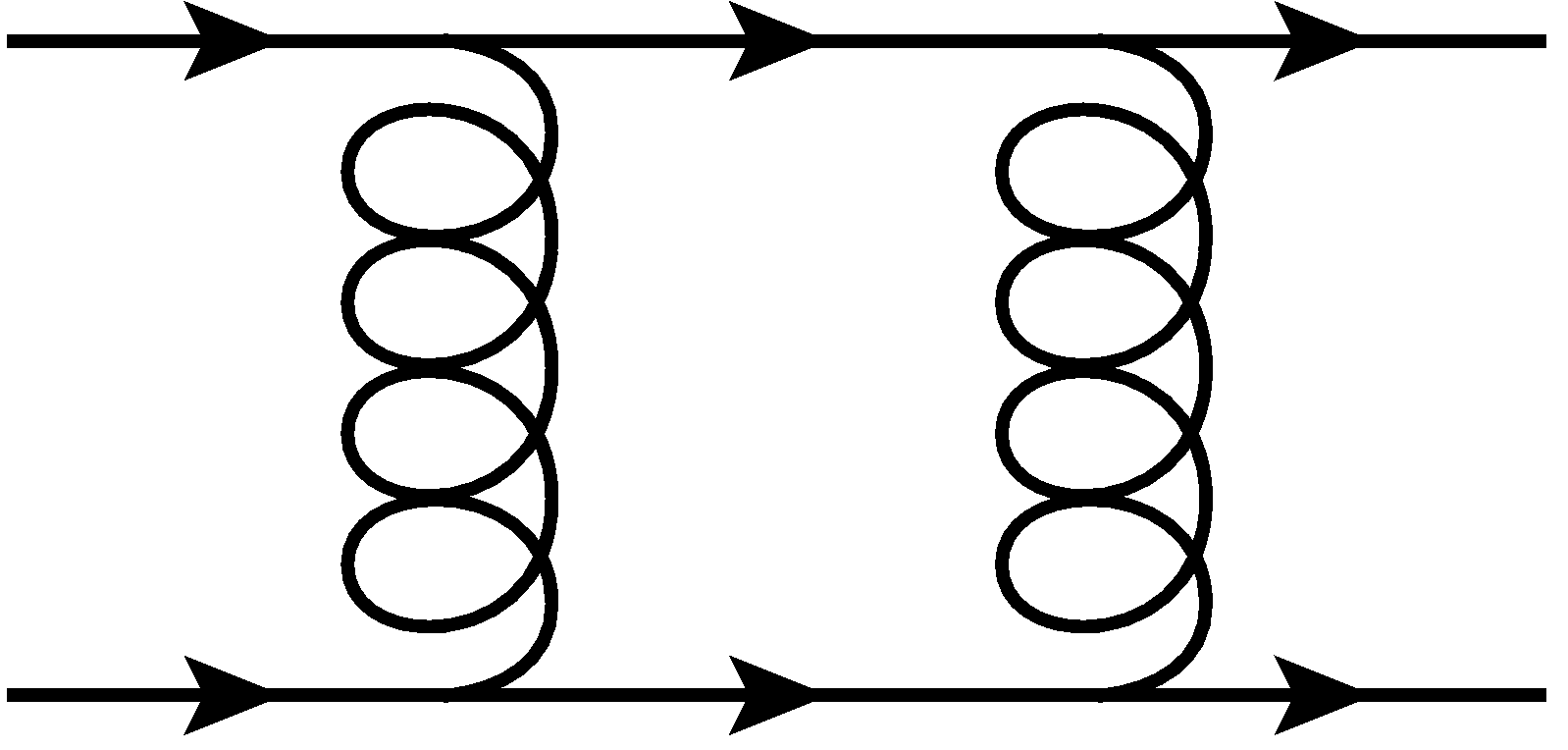} 
    \caption{} 
    \label{fig:box1} 
    \end{subfigure} 
\hspace{.3cm} 
    \begin{subfigure}[t]{0.17\textwidth} 
    \centering 
    \includegraphics[width=\linewidth]{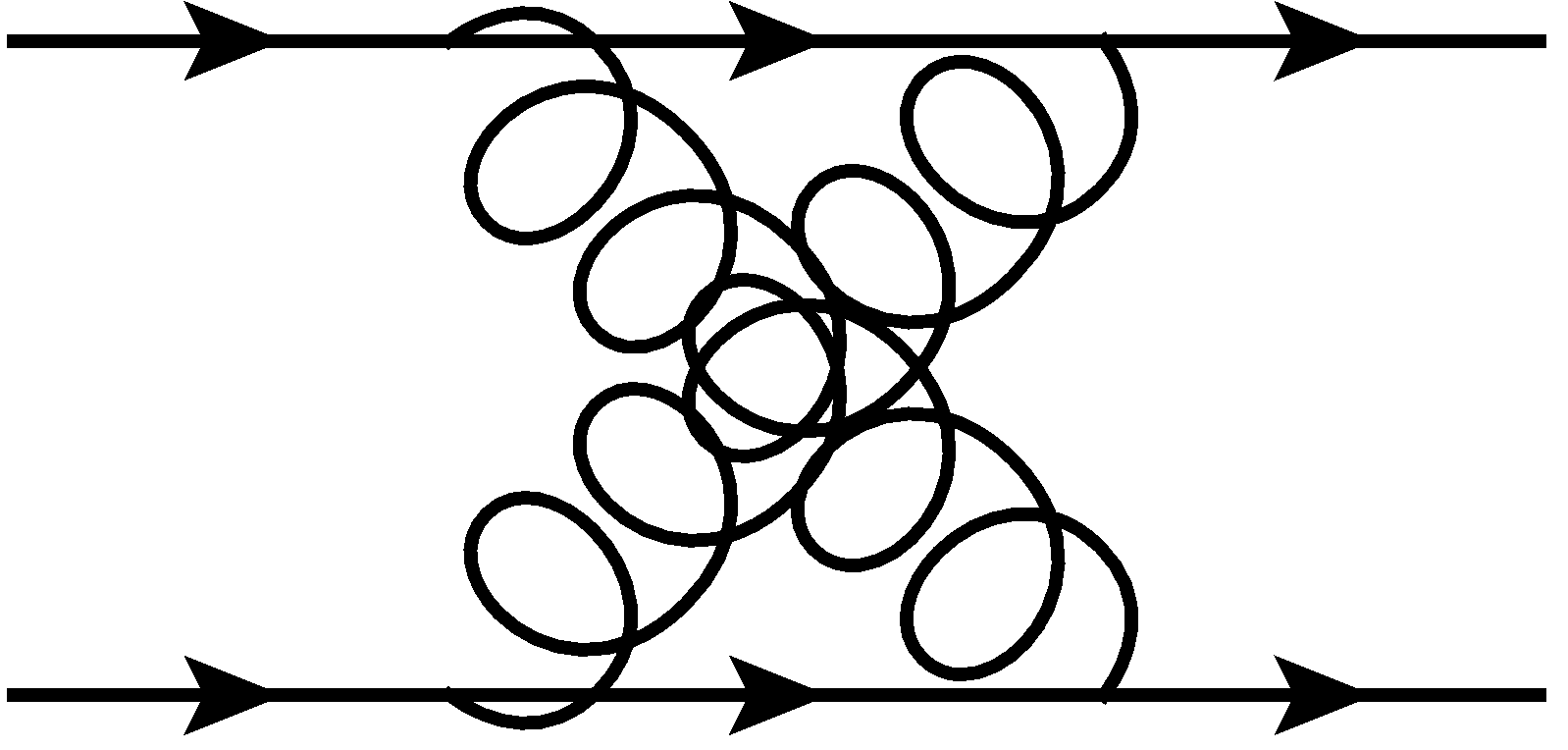} 
    \caption{} 
    \label{fig:box2} 
    \end{subfigure} 
\caption{Sample of graphs contributing to the $2\to2$ forward scattering amplitude. For simplicity, here the two partons are both quarks.} 
\label{fig:graphs} 
\end{figure}

At each perturbative order, the amplitude $M_{ij\to ij}$ contains powers of the high-energy logarithm $L=\log(s/|t|)$, which become large in the forward limit, $s\gg|t|$, and must be resummed to preserve perturbative accuracy.
Such a resummation is far from trivial. 
A useful simplification follows from exploiting the signature symmetry $s \leftrightarrow u$ emerging in this limit, under which the amplitude decomposes into odd and even components, $M_{ij\to ij}^{(-)}$ and $M_{ij\to ij}^{(+)}$.
At leading-logarithmic (LL) accuracy, the odd component has long been known to take the form (see, e.g., Ref.~\cite{Fadin:2020lam})
\begin{align}
    \label{eq:LL_odd}
    &M_{i j \to i j}^{(-)\,\mathrm{LL}} = 
    e^{L\, \omega_G^{\mathrm{LO}}(t,\mu^2)} 
    M_{i j \to i j}^{\mathrm{tree}}(s,t)
\end{align}
where $M_{i j \to i j}^{\mathrm{tree}}$ is the tree-level exchange of Fig.\ref{fig:tree}.
This result can be interpreted as a $t$-channel exchange of a single Reggeon, with the exponent $\omega_G$ being the Regge trajectory of the gluon at lowest-order:
\begin{align}
    \label{eq:regge_LO}
    &\omega_G^{\mathrm{LO}}(t,\mu^2) = 
    C_A \frac{\alpha_s(\mu^2)}{4\pi} 
    \Big(\frac{\mu^2}{-t}\Big)^{\eps}\, 
    \frac{-\Gamma(-\eps)^2 \Gamma(1+\eps)}{\Gamma(-2\eps)}\,.
\end{align}
At next-to-leading logarithmic (NLL) accuracy, the resummation yields\cite{Fadin:2006bj,Fadin:2015zea}
\begin{align}
    \label{eq:NLL_odd}
    &M_{i j \to i j}^{(-)\,\mathrm{NLL}} = 
    C_i^{\mathrm{LO}}(t,\mu^2) e^{L\,  \omega_G^{\mathrm{NLO}}(t,\mu^2)} 
    C_j^{\mathrm{LO}}(t,\mu^2)
    M_{i j \to i j}^{\mathrm{tree}}(s,t)
\end{align}
that generalizes Eq.\eqref{eq:LL_odd} by extending the Regge trajectory up to NLO and by introducing the impact factors $C_{i,j}$ at lowest-order, describing the coupling of the exchanged Reggeon to the external partons.  
The structure of Eq.\eqref{eq:NLL_odd} displays the factorization pattern characteristic of two opposite light-cone directions, with the typical wide rapidity gap encoded in the logarithm $L$. 
The impact factors are analogous to collinear functions, while the Regge exponential correlates the two sectors like a soft contribution, leaving the tree-level amplitude to play the role of the hard function.
Although intriguing, this correspondence breaks down already at NNLL accuracy:
\begin{align}
    \label{eq:NNLL_odd}
    &M_{i j \to i j}^{(-)\,\mathrm{NNLL}} = 
    C_i^{\mathrm{NLO}}(t,\mu^2) e^{L\,  \omega_G^{\mathrm{NNLO}}(t,\mu^2)} 
    C_j^{\mathrm{NLO}}(t,\mu^2)
    M_{i j \to i j}^{\mathrm{tree}}(s,t)
    +
    R_{i j}^{(-)\,\mathrm{LO}}(s,t)
\end{align}
The factorization-breaking contribution $R_{ij}^{(-)}$ is associated with three-Reggeon exchange and hence with a Regge cut\cite{Caron-Huot:2017fxr,Falcioni:2021dgr}. 
It first contributes at NNLL accuracy, starting at NNLO in the fixed-order expansion. 
From this order onwards, separating the single-Reggeon Regge pole from multiple-Reggeon exchange becomes prescription dependent, preventing an unambiguous definition of the Regge trajectory and impact factors. 
This calls for a change of perspective on how forward $2\to 2$ amplitudes factorize.

\section{Reggeon-Glauber correspondence}

The kinematics of forward scattering forces the tree-level exchanged gluon into the Glauber region, characterized by $|l^+ l^-| \ll l_T^2$. 
In other words, single-Reggeon (SR) exchange at lowest order reduces to single-Glauber exchange. 
It is then natural to ask whether this correspondence can be extended to the all-order level:
\begin{align}
    \text{Dressed Single Glauber exchange}
    \quad\leftrightarrow\quad
    \text{Single Reggeon exchange}
\end{align}
where ``dressed'' denotes the inclusion of arbitrary soft and collinear radiative corrections around a single Glauber exchange. 
The left-hand side of this correspondence can be analyzed to all orders using standard Collins-Soper-Sterman factorization methods~\cite{Collins:1989gx,Collins:2011zzd}. Momenta are classified as hard, collinear, or soft according to their scaling, with $\sqrt{s}$ and $\sqrt{|t|}$ setting the relevant hard and soft scales. 
The corresponding regions are disentangled through kinematic approximations and Ward identities. 
Rapidity divergences are regulated by tilting the plus and minus light-cone directions, $n$ and $\nbar$, off the light cone as in Ref.~\cite{Collins:2011zzd}:
\begin{align}
    \label{eq:tilts}
    &n \to (1, e^{-2 y_1}, \vec{0}_T), 
    \qquad
    \nbar \to (e^{2 y_2}, 1, \vec{0}_T)
\end{align}
and the original vectors are recovered in the limits $y_1 \to +\infty$ and $y_2 \to -\infty$, respectively.
The procedure yields
\begin{align}
    \label{eq:SR_factorization}
    &M^{\mathrm{SR}}_{ij \to ij}
    =
    \overline{C}_i(t,\mu^2; L - y_1)
    S_1(t,\mu^2;y_1 - y_2)
    \overline{C}_j(t,\mu^2; y_2 - L)
    M_{i j \to i j}^{\mathrm{tree}}(s,t)
\end{align}
where $\overline{C}_{i,j}$ describe radiation collinear to the external partons and $S_1$ the soft radiation dressing the single-Glauber exchange.
Their rapidity evolution cancels the dependence on the regulators $y_1$ and $y_2$.
We then \emph{define} the Regge trajectory as the rapidity kernel of $S_1$:
\begin{align}
    \label{eq:regge_def}
    \frac{\partial}{\partial y_1} \log{S_1(t,\mu^2; y_1-y_2)} = 
    \omega_G(t,\mu^2)
\end{align}
This definition reproduces the LO and NLO exponents of Eqs.~\eqref{eq:LL_odd} and \eqref{eq:NLL_odd}, but necessarily differs from the cut scheme of Ref.~\cite{Falcioni:2021dgr} starting at NNLO, since it excludes multi-Reggeon contributions.
The impact factors $C_i$ and $C_j$ are recovered by setting $y_1=L$ and $y_2=-L$ in $\overline{C}_i$ and $\overline{C}_j$, respectively, as appropriate for radiation at large positive and negative rapidities.
This perspective provides all-order definitions of the quantities entering single-Reggeon exchange, resolving the ambiguities inherent in its interplay with the multi-Reggeon channel, here excluded from the analysis.

\bigskip

Although powerful, establishing Eq.~\eqref{eq:SR_factorization} is highly nontrivial. 
Glauber exchanges lie at the heart of factorization-breaking mechanisms in gauge theories, obstructing the Ward-identity arguments used to disentangle soft, collinear, and hard subgraphs. Despite major advances in incorporating Glauber interactions into SCET and understanding their role in QCD factorization~\cite{Rothstein:2016bsq,Schwartz:2017nmr,Barcaro:2026dsd,Chen:2026dnj}, no fully general framework valid at arbitrary accuracy is currently available.
Here, we consider the peculiar case in which the number of Glauber gluons is kept \textbf{fixed} and their Glauber scaling is imposed by the process kinematics, rather than via pinch singularities in loop momentum space. 
Soft gluons are decoupled from collinear subgraphs by retaining only the lightcone component that couples to the large component of the collinear momentum, and replacing soft-collinear interactions with eikonal couplings.
Ward identities complete the factorization, reorganizing color factors and removing all dependence on soft momenta from the collinear subgraphs. 
These arguments remain valid in the presence of a Glauber insertion, provided it is consistently retained throughout the derivation.
A sketch of the proof follows by considering two soft gluons of momenta $k_1$ and $k_2$ attached to a collinear-to-plus fermion line and dressing a single Glauber insertion of momentum $l$. The line carrying $l$ and all momenta to its left flow into the collinear subgraph, while those to its right flow out. After some color algebra, we obtain:
\begin{align}
    &\mathcal{C}_q^{\alpha\beta_1 \beta_2, a b_1 b_2}(p_1,p_3,k_1,k_2) \mathcal{S}_{\alpha\beta_1 \beta_2}^{a b_1 b_2}(k_1,k_2,l) = 
    \notag \\
    &\;=
    \Big[ 
    \begin{gathered}
    \includegraphics[height=1.1cm]{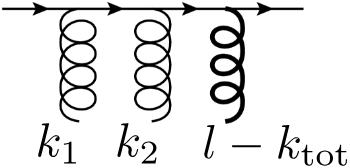}
    \end{gathered}
    +
    \begin{gathered}
    \includegraphics[height=1.1cm]{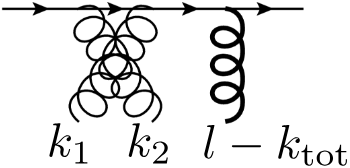}
    \end{gathered}
    +
    \begin{gathered}
    \includegraphics[height=1.1cm]{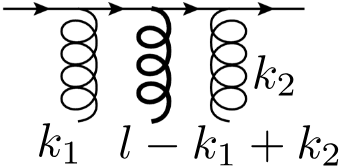}
    \end{gathered}
    +
    \text{ permutations}
    \Big]\,\mathcal{S}_{\alpha\beta_1 \beta_2}^{a b_1 b_2}(k_1,k_2,l)
    \notag \\
    &\;=
    \mathcal{C}_q^{+\beta_1 \beta_2, a b_1 b_2}(p_1,p_3,\widehat{k}_1,\widehat{k}_2) \frac{\widehat{k}_{1\beta_1} n_\rho}{k_1 \cdot n + i\,0}
    \frac{\widehat{k}_{2\beta_2} n_\sigma}{k_2 \cdot n + i\,0}
    \mathcal{S}^{- \rho \sigma,a b_1 b_2}(k_1,k_2,l)
    \notag \\
    &\;=
    [\ubar_3 (-i g \gamma^+) u_1]
    \Big\{
    t^a \frac{-i g n^{\beta_1} t^{b_1}}{k_1 \cdot n + i\,0} \frac{-i g n^{\beta_2} t^{b_2}}{k_{\mathrm{tot}} \cdot n + i\,0}
    +
    t^a 
    \frac{-i g n^{\beta_2} t^{b_2}}{k_2 \cdot n + i\,0} \frac{-i g n^{\beta_1} t^{b_1}}{k_{\mathrm{tot}} \cdot n + i\,0}
    \notag \\
    &\quad
    +
    \frac{-i g n^{\beta_1} t^{b_1}}{k_1 \cdot n + i\,0} \frac{-i g n^{\beta_2} t^{b_2}}{k_{\mathrm{tot}} \cdot n + i\,0} t^a
    +
    \frac{-i g n^{\beta_2} t^{b_2}}{k_2 \cdot n + i\,0} \frac{-i g n^{\beta_1} t^{b_1}}{k_{\mathrm{tot}} \cdot n + i\,0} t^a
    \notag \\
    &\quad
    +
    \frac{-i g n^{\beta_1} t^{b_1}}{k_1 \cdot n + i\,0} 
    t^a 
    \frac{-i g n^{\beta_2} t^{b_2}}{k_2 \cdot n + i\,0}
    +
    \frac{-i g n^{\beta_2} t^{b_2}}{k_2 \cdot n + i\,0}
    t^a
    \frac{-i g n^{\beta_1} t^{b_1}}{k_1 \cdot n + i\,0} 
    \Big\}\mathcal{S}^{- \rho \sigma,a b_1 b_2}(k_1,k_2,l) + \pst
\end{align}
This has the structure of the tree level amplitude multiplying a soft factor containing a Wilson line running from the far past towards the far future along the direction $n$, with a colored insertion at the point where the Glauber gluon attaches. 
The argument above does not determine the exact form of this insertion, which will be presented in a forthcoming paper. For present purposes, we introduce the following preliminary coordinate-space operator:
\begin{align}
    \label{eq:S1_def_naive}
    &\mathbf{S}_1(z_\perp)
    \sim
    W_{z_\perp \to \infty, \,n_1} A_+(z_\perp) W_{-\infty \to z_\perp, \,n_1}
    \otimes
    W_{0 \to \infty, \,n_2} A_-(0) W_{-\infty \to 0, \,n_2}
\end{align}
with the usual path-ordered definition
\begin{align}
    \label{eq:WL_def}
    &W_{x \to y, \,\gamma} = \mathcal{P}\mathrm{exp}\big\{
    -i g \int_x^y d\gamma^\mu A_\mu(\gamma) 
     \big\}\,.
\end{align}
The two Wilson lines are directed along the tilted directions of Eq.~\eqref{eq:tilts} and separated by the transverse displacement $z_\perp=(0,0,\vec z_T)$.
The soft function in Eq.~\eqref{eq:SR_factorization} will then be defined as the vacuum expectation value of a suitable gauge-invariant completion of this operator, projected onto the color-octet channel and Fourier transformed to momentum space.

\section{Bridging Regge trajectory and Collins-Soper kernel}

\begin{figure}[t] 
\centering 
    \begin{subfigure}[t]{.35\textwidth} 
    \centering 
    \includegraphics[width=\linewidth]{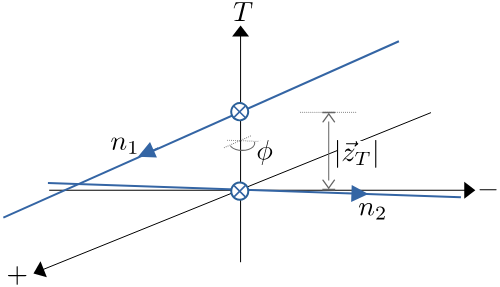} 
    \caption{} 
    \label{fig:soft_forw} 
    \end{subfigure} 
\hspace{1cm} 
    \begin{subfigure}[t]{.35\textwidth} 
    \centering 
    \includegraphics[width=\linewidth]{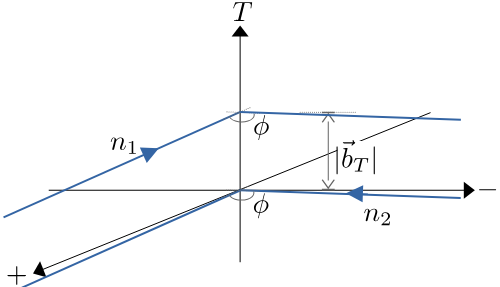} 
    \caption{} 
    \label{fig:soft_tmd} 
    \end{subfigure} 
\caption{Soft operators for (a) single-Reggeon exchange and (b) TMD factorization. The cusp angles and transverse separations are indicated, together with the orientation and directions of the Wilson lines.} 
\label{fig:soft_factors} 
\end{figure}

The preliminary soft operator of Eq.~\eqref{eq:S1_def_naive} describes two infinite Wilson lines running close to the light cone, equipped with source insertions and separated by a transverse distance $|\vec z_T|$. 
Apart from the details of the insertions, its essential properties will remain unchanged in the final formulation.
The geometry of this configuration bears a close resemblance to the structure arising in transverse-momentum-dependent (TMD) factorization, where the soft factor is a staple-like Wilson loop closing at infinity, see Fig.\ref{fig:soft_factors}.
The similarity becomes particularly evident at the level of UV renormalization: in the lightcone limit, it is insensitive to many details of the operator and depends only on its cusp structure. 
In particular, the TMD soft operator contains two cusps, whereas the single-Reggeon operator involves one effective cusp generated by the instantaneous Glauber interaction between the source insertions.
This yields the comparison
\begin{align}
    \label{eq:UV_ren}
    &\frac{\partial}{\partial \log{\mu}} \log{\mathbf{S}_1(z_\perp)} = 
    \phi\,\Gamma_K(\alpha_s) + \dots
    \qquad
    \frac{\partial}{\partial \log{\mu}} \log{\mathbf{S}_\mathrm{tmd}(b_\perp)} = 
    -2 
    \phi\,\Gamma_K(\alpha_s) + \dots
\end{align}
where $\phi=y_1-y_2$ is the cusp angle and $\Gamma_K$ is the universal lightlike cusp anomalous dimension governing the asymptotic linear growth at large $\phi$, currently known up to four loops (see Ref.~\cite{Grozin:2022umo} for a review). The dots denote subleading terms in this limit, while the relative minus sign follows from the opposite orientations of the overlapping Wilson-line segments.
Remarkably, the UV renormalization of $\mathbf{S}_1$ automatically yields $\Gamma_K$ as the anomalous dimension of the Regge trajectory defined in Eq.\eqref{eq:regge_def}, providing an all-order argument for a conjecture previously verified through three loops\cite{Falcioni:2021buo}.
The comparison extends to rapidity evolution and it highlights a deep analogy between the rapidity kernels of the two operators. 
The TMD soft factor satisfies
\begin{align}
    \label{eq:CSk_def}
    \frac{\partial}{\partial y_1} \log{S_{\mathrm{tmd}}(b_T^2,\mu^2; y_1-y_2)} = 
    K_G(b_T^2,\mu^2)
\end{align}
where $K_G$ is the Collins-Soper kernel~\cite{Collins:1981uk} in the adjoint representation, currently known up to N$^3$LO\cite{Moult:2022xzt,Duhr:2022yyp}. 
Its anomalous dimension is $-2\Gamma_K$, implying that the combination
\begin{align}
    \label{eq:discrepancy}
    &\Delta(t,\mu^2) =
    \omega_G(t,\mu^2) + \frac{1}{2}K_G(t,\mu^2)
\end{align}
is renormalization group (RG) invariant. 
More specifically, the function $\Delta$ quantifies the difference between the Regge trajectory and the Collins-Soper kernel. 
It vanishes at LO and remains remarkably simple at higher orders because of extensive cancellations. 
At NLO, only a common term proportional to $\zeta_3$ survives:
\begin{align}
    &\Delta^{[1]} = \Big(\frac{\alpha_s}{4\pi}\Big)^2 \, C_A^2 12 \zeta_3
\end{align}
Beyond NLO, the comparison is complicated by the prescription dependence of the Regge trajectory. Nevertheless, using the definition of Ref.~\cite{Falcioni:2021dgr}, one finds
\begin{align}
    &\Delta^{[2]} = -\Big(\frac{\alpha_s}{4\pi}\Big)^3 \, C_A^3 \Big[
    \frac{4}{27}\big(125-304 \frac{n_f}{N} + 44 \frac{n_f^2}{N^2}\big) \zeta_3 
    +
    16 \zeta_2 \zeta_3 + 80 \zeta_5
    \Big]
\end{align}
a substantial simplification compared with the original expressions, which contain up to fifteen distinct terms.

\section{Conclusions}

In this work, we have laid the foundations for an all-order factorization of single-Reggeon exchange and provided a preliminary construction of the associated soft operator. 
This framework builds a bridge between the Regge limit of amplitudes and TMD factorization, leading to the RG-invariant quantity $\Delta$, which measures the difference between the Regge trajectory and the Collins-Soper kernel. 
The remarkable simplicity of its perturbative expansion points to a deeper geometric relation between the two rapidity kernels.
This framework paves the way to exploit this connection to determine the Regge trajectory from the Collins-Soper kernel. Since the latter can be extracted from calculations requiring one loop order less, this approach could provide a substantially more efficient route to high-order predictions. Ultimately, extending the formalism to multiple Glauber exchanges will be essential to describe multi-Reggeon dynamics and advance high-energy factorization beyond leading-logarithmic accuracy.

\section*{Acknowledgements}

The participation in this workshop was partially funded by COST Action CA24159 through a Young Researcher and Innovator Conference Grant (E-COST-GRANT-CA24159-be87c4a7).

\end{document}